\documentclass[apj,twocolumn]{openjournal}
\usepackage{natbib} 
\usepackage[dvipsnames]{xcolor}
\usepackage{aas_macros} 
\usepackage{amssymb}
\usepackage{amsmath}
\usepackage[title]{appendix}
\usepackage{hyperref}	
\hypersetup{colorlinks=true,linkcolor=blue,citecolor=blue,filecolor=blue,urlcolor=blue}
\usepackage[caption=false]{subfig}
\usepackage{soul}                          
\usepackage{xspace}
\usepackage{booktabs}
\usepackage{savesym}
\savesymbol{tablenum}
\usepackage{siunitx}
\usepackage{pifont}
\restoresymbol{SIX}{tablenum}

\DeclareSIUnit\Msun{\ensuremath{M_\odot}}
\DeclareSIUnit\Zsun{\ensuremath{Z_\odot}}
\DeclareSIUnit\hred{\ensuremath{\textit{h}}}

\begin{document}
\title[Turbulent H~{\small II} Regions]{Inferring Interstellar Medium Density, Temperature, and Metallicity from Turbulent H~{\small II} Regions\vspace{-15mm}}
\author{Larrance Xing$^{1,*}$, Nicholas Choustikov$^{2}$, 
Harley Katz$^{1,3,*}$, \& Alex J. Cameron$^{4,5,2}$}
\thanks{$^*$E-mail: \href{mailto:larrance@uchicago.edu}{larrance@uchicago.edu}}
\thanks{$^*$E-mail: \href{mailto:harleykatz@uchicago.edu}{harleykatz@uchicago.edu}}

\affiliation{$^{1}$Department of Astronomy \& Astrophysics, University of Chicago, 5640 S Ellis Avenue, Chicago, IL 60637, USA}
\affiliation{$^{2}$Sub-department of Astrophysics, University of Oxford, Keble Road, Oxford OX1 3RH, United Kingdom}
\affiliation{$^{3}$Kavli Institute for Cosmological Physics, University of Chicago, Chicago IL 60637, USA}
\affiliation{$^{4}$Cosmic Dawn Center (DAWN), Copenhagen, Denmark}
\affiliation{$^{5}$Niels Bohr Institute, University of Copenhagen, Jagtvej 128, DK-2200, Copenhagen, Denmark}

\begin{abstract}
Reliable nebular emission line diagnostics are essential for accurately inferring the physical properties (e.g. electron temperature, density, pressure, and metallicity) of H~{\small II} regions from spectra. When interpreting spectra, it is typical to adopt a single zone model, e.g. at fixed density, pressure, or temperature, to infer H~{\small II} region properties. However, such an assumption may not fully capture the complexities of a turbulent interstellar medium. To understand how a complex density field driven by supersonic turbulence impacts nebular emission lines, we simulate 3D H~{\small II} regions surrounding a single O star, both with and without supersonic turbulence. We find that turbulence directly impacts the values of common strong line ratios. For example turbulent H~{\small II} regions exhibit systematically higher [N~{\small II}]/H$\alpha$, lower [O~{\small III}]/H$\beta$, and lower O32, compared to homogeneous H~{\small II} regions with the same mean density and ionizing source. These biases can impact inferences of metallicity, ionization parameter, excitation, and ionization source. For our choice of turbulence, direct $T_e$ method metallicity inferences are biased low, by up to 0.1~dex, which is important for metallicity studies, but not enough to explain the abundance discrepancy problem. Finally, we show that large differences between measured electron densities emerge between infrared, optical, and UV density indicators. Our results motivate the need for large grids of turbulent H~{\small II} regions models that span the range of conditions seen at both high and low redshift to better interpret observed spectra.
\end{abstract}
\keywords{ISM, galaxy formation}

\section{Introduction}
\label{sec:introduction}

Nebular emission line diagnostics represent a powerful probe of the physical properties of the interstellar medium (ISM). The electron temperature, density, pressure, and metallicity of H~{\small II} regions can all be probed by various combinations of emission lines that originate in ionized gas \citep{Kewley:2019,Kewley:2019b}. Likewise, diagnostics from low-ionization state atoms and molecules can be used to trace similar properties in neutral or diffuse ionized gas.

The physical conditions (e.g. electron temperature, electron number density, and ion number density) of the ISM strongly impact the emission from H~{\small II} regions. For example, an ion can be used to trace electron temperature in an H~{\small II} region if the species is both abundant enough and the first two excited states are at low enough energy so that they can be sufficiently populated by collisional excitation. An example of such an ion is O$^{++}$. O$^{++}$ has an ionization energy of $35.1$~eV which is low enough to be photoionized by stars, while the next ionization state requires 54.9~eV, which is typically higher than that of photons emitted by a massive, metal-enriched O-star --- hence the ion is sufficiently abundant. Next, the first two excited state (from the $^3$P ground state) are $^1$D and $^1$S, which require $\sim2.5$~eV and $\sim5.3$~eV, respectively to populate. Assuming the electron velocity distribution function is well modeled by a Maxwell-Boltzmann distribution, which depends only on temperature, the ratio of emission from these two states (i.e. the 5007~\AA\ and 4363~\AA\ lines) probe the electron temperature. Other ions with structure suitable for tracing electron temperature that are commonly observed in H~{\small II} regions include N$^+$, O$^+$, S$^{++}$, etc. Similarly, electron density can be probed, for example, by ions that have closely-spaced doublet states (e.g. O$^+$, S$^{+}$, Ar$^{+++}$). Because the doublet state requires approximately the same excitation energy, the difference in emissivity is dictated by differences in the Einstein coefficient for spontaneous de-excitation and collisional de-excitation. The former is measured from quantum mechanical calculations and the latter represents the probe of electron density. The simplicity of this physics makes such calculations readily applicable to observed galaxy spectra with open-source software \citep[e.g.][]{Dere1997,Luridiana2015}.

A key assumption about these calculations is that all of the emission originates from a single ``zone'' that has a uniform density, temperature, metallicity, etc. This assumption is untrue both within individual H~{\small II} regions \citep[e.g.][]{Rubin1968} as well as across a galaxy as a whole \citep[e.g.][]{Kobulnicky1999,Harikane2025}. To address this issue, there are two approaches. First, numerous codes have been developed that statistically apply multi-zone models to observed spectra to infer galaxy and H~{\small II} region properties \citep[e.g.][]{Lebouteiller2022,Marconi2024}. For such inference, either the distribution function of H~{\small II} region properties must be parametrized or for non-parametric models, a cap or regularization must be applied to limit the number of zones. Alternatively, one can attempt to forward model the detailed structure of individual H~{\small II} regions or a whole galaxy (comprised of multiple H~{\small II} regions. While modeling entire galaxies remains difficult, public codes such as CLOUDY \citep{Ferland1998,chatzikos20232023releasecloudy} or MAPPINGS \citep{1996ApJS..102..161D,Sutherland2018} have made H~{\small II} region models widely accessible and they have been extremely successful in being able to reproduce the distribution of observed galaxy and H~{\small II} region properties seen in large surveys \citep[e.g.][]{Kewley2001}.

However, photoionization models of H~{\small II} regions have some key limitations. First, they often assume simplified geometries (e.g. spherical or plane-parallel) and either constant density or pressure\footnote{Note that this is not true of all photoionization models in the literature but this is anecdotally a reasonable description of what is generally assumed.}. It is now well established that the geometry of an H~{\small II} region can have significant effects on the resultant emission \citep[e.g.][]{Jin:2022b}. Second, typical photoionization models assume that the cloud is in equilibrium (although c.f. \citealt{Gnat2007,Kumar2025} in the context of the circumgalactic medium --- CGM). However, this assumption may break down in the case where the timescale for the ionization front to reach the Strömgren radius becomes shorter than the main-sequence lifetime of massive stars or if the cooling time is shorter than the recombination time (see e.g. \citealt{Richings2022,Katz:2022a} for discussion of the latter). Finally, the ISM is known to be turbulent and when the turbulence is supersonic, emission line ratios can differ from a static medium \citep[e.g.][]{Gray:2017}. 

The primary reason more complicated models of H~{\small II} regions and galaxies have been less explored is that they are computationally much more expensive. While grids of millions of, for example, CLOUDY models can be readily computed on modern supercomputer architectures, the same is not true for 3D H~{\small II} region models. Nevertheless, there are now numerous codes that can address the key issues with photoionization models \citep[e.g.][]{Gray:2015,Ziegler2018,Jin:2022a,Katz:2022b,Smith2023,Chan2025,McClymont2025}, with a subset able to couple the detailed non-equilibrium chemistry to the hydrodynamics. 

The purpose of this work is to consider one step beyond a simplistic H~{\small II} region model and quantify how emission line diagnostics respond to a non-trivial H~{\small II} region geometry. We focus on line-ratio diagnostics for metallicity and electron density. In our idealized experiments, we first compute the emission from the H~{\small II} region around a single O~star embedded in a uniform density medium. Next, we drive supersonic turbulence around the star, recompute the emission, and compare the resultant line ratios to the homogeneous case. This allows us to estimate the impact of an inhomogeneous density structure on nebular emission line diagnostics.

This paper is arranged as follows. We describe the numerical simulations and the methods for measuring metallicity and density in Section \ref{sec:methods}. We present our results in Section \ref{sec:results}, and discuss how our results compare with previous studies pertaining to line ratio diagnostics in Section \ref{sec:discussion}. 

\section{Numerical Simulations}
\label{sec:methods}

All simulations in this work are run with the {\small RAMSES-RTZ} code \citep{Katz:2022b} and the PRISM ISM model \citep{Katz:2022a}, which is a fork of {\small RAMSES} \citep{Teyssier2002} and {\small RAMSES-RT} \citep{Rosdahl2013,Rosdahl2015}. {\small RAMSES-RTZ} has been benchmarked directly against {\small CLOUDY} in numerous ways and shown to reproduce the emission line luminosities of constant-density H~{\small II} regions to better than 10\% on average \citep{Katz:2022b,Katz2024}, which is consistent with the differences seen between {\small CLOUDY} and {\small MAPPINGS} \citep{Agostino2019}. We consider two types of simulations: constant density (homogeneous) models and turbulent boxes. Our numerical setup shares many similarities with the work of \cite{Jin:2022b}, with a key difference being that we drive turbulence through the box to obtain a complex density distribution rather than adopting a fractal geometry. 

\subsection{Calibrations with Homogeneous Boxes}

Before we can study the impact of a non-trivial density distribution on emission line ratios, it is crucial to calibrate our emission line diagnostics predicted by {\small RAMSES-RTZ} for a constant density medium. To this end, we begin by placing a single O4V star, modeled as a blackbody with a temperature of 42,900~K\footnote{Note that by modeling our central source as a blackbody, we are over-predicting the number of ionizing photons of a typical O star with the same effective surface temperature. Nevertheless, these experiments are designed to be demonstrative of the qualitative trends for how turbulence impacts H~II region emission lines.}, at the center of a computational domain with side length 10~pc, with a constant density. The source emits $1.25\times10^{50}$ ionizing photons per second. Our homogeneous models range from 10$-$10,000 H~${\rm cm^{-3}}$ in density. The domain is split into $256^3$ elements and the box size is varied depending on density so that the H~{\small II} region is ionization-bounded. The simulation is then evolved until the ionization front reaches the Strömgren radius and the system reaches steady-state. Note that here we do not model the hydrodynamics of the system and only consider the impact of photoionization and photoheating. In other words, the gas distribution remains static (ignoring radiation pressure and the over-pressurization of the H~{\small II} region from photoheating). Hence, our setup is as similar as possible to standard photoionization models with the only difference being the 3D geometry. The radiation is modeled in eight frequency bins following \cite{Kimm2017,Katz2024}. We consider five different metallicities, 0.01, 0.05, 0.1, 0.5, and 1 $Z_{\odot}$ with metal abundance patterns adopted from \cite{Asplund2009} and scaled by metallicity.

Emission lines are computed from the simulation on a cell-by-cell basis using the electron temperature, density, and ionization states predicted by {\small RAMSES-RTZ}. For UV and optical emission lines, we use {\small PyNeb} \citep{Luridiana2015} to compute emissivities and adopt atomic data from {\small CHIANTI} Version 10.1 \citep{2023ApJS..268...52D} where possible. Otherwise we adopt the default atomic data in {\small PyNeb}. Each collisionally excited line is modeled with 15 energy levels. We also consider mid-IR and far-IR emission lines. These are computed as in {\small RAMSES-RTZ} for cooling with a method that closely follows {\small CLOUDY} \citep{Ferland:2017}.

A full list of all diagnostics used in this work for electron density and temperature is provided in Table \ref{tab:diagnostics}, while the method of inferring metallicities is described in Section \ref{sec:calib_metal}.

\begin{table}
	\centering
	\caption{Emission line diagnostics used in this work.}
	\label{tab:diagnostics}
	\begin{tabular}{llcr} 
        \hline
        Name & Line Ratio & Diagnostic \\
        \hline
        ${R}_{{\rm O~II}}$   & ${\rm [O~II]}~\lambda3728 \;/\; {\rm [O~II]}~\lambda3726$  & $n_e$  \\
        ${R}_{{\rm O~III}}$   & ${\rm [O~III]}~88\mu  {\rm m} \;/\; {\rm [O~III]}~52\mu  {\rm m}$ & "  \\
        ${R}_{{\rm S~II}}$   & ${\rm [S~II]}~\lambda6717 \;/\; {\rm [S~II]}~\lambda6731$ & "  \\
        ${R}_{{\rm S~III}}$   & ${\rm [S~III]}~33\mu  {\rm m} \;/\; {\rm [S~III]}~18\mu  {\rm m}$  & "   \\
        ${R}_{{\rm C~III}}$   & ${\rm [C~III]}~\lambda1906 \;/\; {\rm C~III]}~\lambda1908$ & "   \\
        ${R}_{{\rm N~II}}$   & ${\rm [N~II]}~205\mu {\rm m} \;/\; {\rm [N~II]}~122\mu  {\rm m}$ & "   \\
        \hline
        ${R}_{\lambda4363}$ & ${\rm [O~III]} \lambda4363 \;/\; {\rm [O~III]}~\lambda5007$ & $T_e$  \\
        ${R}_{\lambda\lambda3727}$ & ${\rm [O~II]}~\lambda\lambda3727\; /\; {\rm [O~II]}~\lambda\lambda\lambda\lambda 7320,7330$ & " \\
        ${R}_{\lambda3726}$ & ${\rm [O~II]}~\lambda3726\; /\; {\rm [O~II]}~ \lambda\lambda\lambda\lambda7320,7330$ & " \\
        \hline
	\end{tabular}
\end{table}

\subsubsection{Density}
\label{sec:calib_den}

\begin{figure*}
    \includegraphics[width=\textwidth]{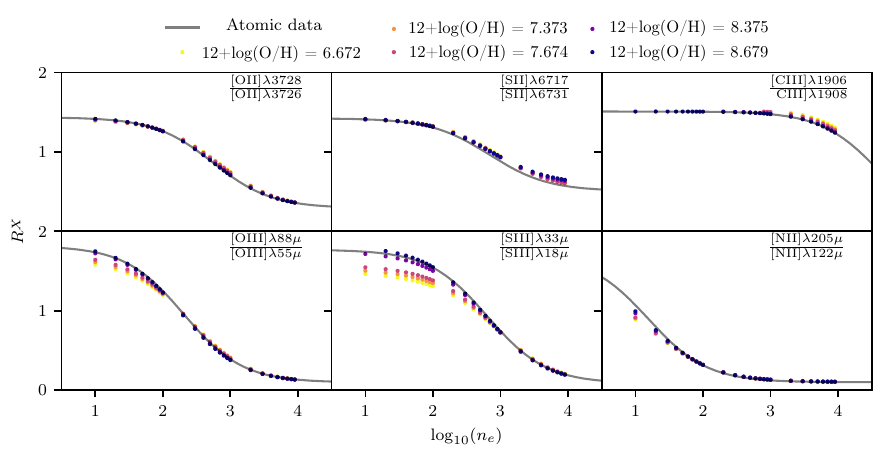}
   \caption{Normalized emission line ratio diagnostics as a function of electron density for each of the homogeneous boxes coloured by their metallicity. The smooth curves show the predicted ratio for a uniform electron temperature of 10$^4$~K.}
    \label{fig:density_cal}
\end{figure*}

As described, electron density can be inferred from closely-spaced doublet states that have different critical densities. We show in Figure~\ref{fig:density_cal} six line ratios measured from the homogeneous simulations as a function of density for the five metallicites. For comparison, we show the theoretical values for these line ratios as computed with {\small PyNeb} assuming a temperature of $10^4$~K. In general, we find that these diagnostics follow the expected relations. Furthermore, we confirm that the curves traced by ratios of IR lines show an added metallicity dependence as predicted by \cite{Kewley:2019}, with lower metallicity realizations producing a decreased ratio at lower electron number densities.

\subsubsection{Temperature}
\label{sec:calib_tem}

Unlike density\footnote{In actuality, electron density is not exactly constant within the Strömgren sphere, particularly at the transition regions at the edge. This impacts ions like S~II that can exist in both ionized and neutral gas more than others. Nevertheless, our models can be reasonably well approximated as having constant electron density within $R_{\rm S}$.}, temperature is not constant within the Strömgren sphere. Models with different metallicites can exhibit temperature gradients, either positive or negative, depending on metallicity \citep[e.g.][]{1992AJ....103.1330G,Stasinska2005}. For this reason, when referring to temperature, we adopt the ``line'' temperature, defined by \cite{Stasinska:1978} as:
\begin{equation}
\label{eq:line_temp}
\epsilon_{\lambda}(T_{\rm line}) = \frac{\int n_X n_e \epsilon_{\lambda}(T) dV}{\int n_X n_e dV},
\end{equation}
where $n_X$ is the number density of the ionic species. $T_{\rm line}$ represents the temperature which when evaluating the emissivity of a line gives the volume-averaged emissivity weighed by the electron and ion number densities. This temperature is useful because when used to calculate metallicity with a ``direct'' method, it gives the correct answer \citep[e.g.][]{Cameron:2023}. When temperature is constant, the temperature measured from emission line ratios is the line temperature; however, if a temperature distribution is present, $T_{\rm line}$ is not necessarily that measured from an emission line ratio. 

\begin{figure}
    \includegraphics{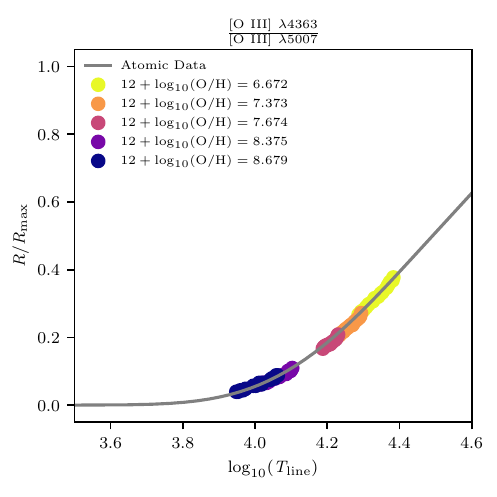}
    \caption{Normalized $R_{\lambda 4363}$ metallicity diagnostic as a function of line temperature (defined by Equation \ref{eq:line_temp}) for each of the homogeneous boxes colored by their metallicity. The curve shows the predicted ratio for a uniform electron number density of $\SI{300}{\per\centi\metre\cubed}$.}
    \label{fig:temp_cal}
\end{figure}

We show in Figure \ref{fig:temp_cal} the electron temperature as measured from the $R_{\lambda 4363}$ ratio for each of five metallicities and various densities as a function of $T_{\rm line}$. For guidance, we show theoretical values computed with atomic data produced from {\small PyNeb}, assuming an electron density of $\SI{300}{\per\centi\metre\cubed}$. Overall, within the homogeneous H~{\small II} regions, the temperature diagnostic works reasonably well despite small variations in temperature.  The ``ratio temperature'' well-matches the line temperature for [O~{\small III}]~$\lambda$4363/$\lambda$5007, which is key for the metallicity estimates we discuss below.

\subsubsection{Metallicity}
\label{sec:calib_metal}
Oxygen abundance (metallicity) is calculated using the direct $T_e$ method. We use the ratio of auroral lines to other forbidden collisionally excited lines to measure the electron temperature (ratio temperature). This temperature is then used to infer the emissivities of the observed emission lines. The oxygen abundance is estimated as 
\begin{equation}
\label{eq:metallicity}
\begin{split}
\frac{\mathrm{O}}{\mathrm{H}} &\approx 
\frac{\mathrm{O}^+}{\mathrm{H}^+} +
\frac{\mathrm{O}^{++}}{\mathrm{H}^+} \\[3pt]
&= 
\frac{L_{\lambda\lambda3727}}{L_{\mathrm{H}\beta}}
\frac{\epsilon_{\mathrm{H}\beta}(T)}{\epsilon_{\lambda\lambda3727}(T)} +
\frac{L_{\lambda5007}}{L_{\mathrm{H}\beta}}
\frac{\epsilon_{\mathrm{H}\beta}(T)}{\epsilon_{\lambda5007}(T)} .
\end{split}
\end{equation}
This equation is a relatively good approximation because the first ionization states of O and H are strongly coupled due to charge exchange reactions and as a result of the high energy required to ionize O to O$^{3+}$, the abundance of O$^{3+}$ is negligible in stellar-dominated H~{\small II} regions to an extent where it can be safely ignored for metallicity calculations \citep{2021ApJ...922..170B}. We note that the temperatures that enter the emissivity calculation need not be the same. This is discussed further below. 

\subsection{Turbulent Boxes}
Having confirmed that our constant density models reproduce the expected behaviour between emission line luminosity and temperature/density, here we consider the impact of a turbulent density field on the same emission line diagnostics. The numerical setup is nearly identical to the homogeneous simulations apart from the density distribution. 

To generate turbulent initial conditions, we start with the homogeneous volume with density of 300~H~${\rm cm^{-3}}$ in a 10~pc box with spatial resolution of 0.039~pc. We then drive turbulence with a mach number of 5.5 and a natural mixture of compressive to solenoidal modes (see e.g. \citealt{Federrath2010}). 
The turbulence is modeled using a Ornstein-Uhlenbeck process \citep[e.g.][]{Eswaran1988,Federrath2010} as implemented in {\small RAMSES} \citep[see details in][]{Brucy2024}. We adopt a parabolic forcing spectrum for the turbulence which is injected on scales of $1L_{\rm box}-\frac{1}{3}L_{\rm box}$. We first let the turbulence develop for a few autocorrelation timescales and then we output the density field every autocorrelation timescale for ten cycles. This provides us ten realizations of a turbulent density field for each metallicity. For each output, we freeze the density field, inject the same O-star at the center of the box, and evolve the radiation field until the simulation reaches a steady-state. We consider metallicities between 5\% and 100\% of solar (i.e. excluding the lowest metallicity used in the homogeneous simulations). Because of the finite box size and the complex density field that is well represented by a lognormal PDF, the escape fraction of ionizing photons is $>0$ in some realizations, but typically remains below 10\%.

\begin{figure*}
\centering
	\includegraphics[width=0.9\textwidth]{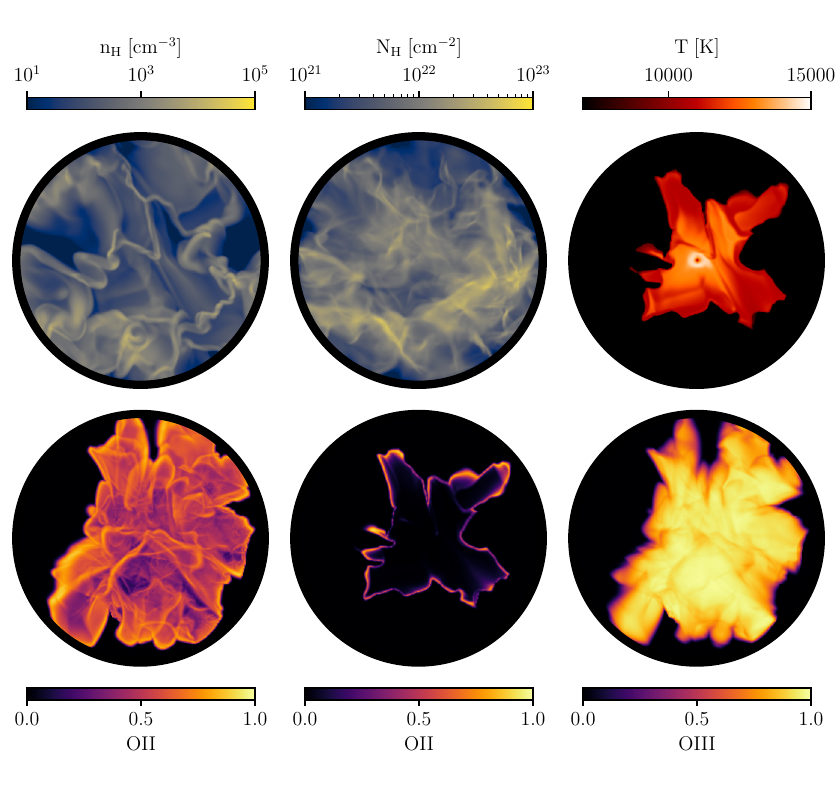}
   \caption{Top row, from left to right: hydrogen density slice, hydrogen column density map, temperature slice. Bottom row, from left to right: integrated O~{II} map, O~{II} slice, integrated O~{III} map. The diameter of each panel is 10~pc.}
    \label{fig:turb}
\end{figure*}

\begin{figure*}
    \centering
	\includegraphics[width=0.9\textwidth]{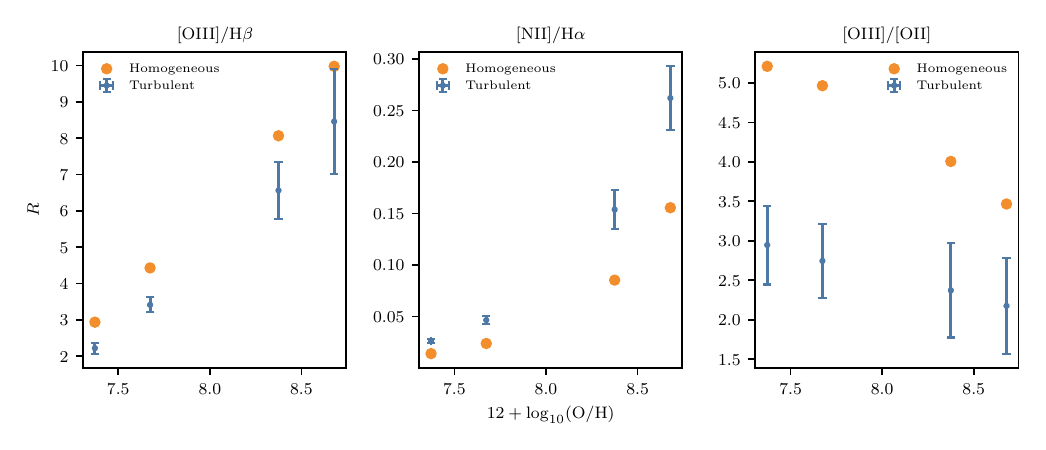}
    \caption{Comparison of diagnostic emission-line ratios from the homogeneous and turbulent simulations as a function of metallicity. [O~III]/H$\rm\beta$, [N~II]/H$\rm\alpha$, and [O~III]/[O~II] are showed in the left, center, and right, respectively. Orange points represent data from homogeneous simulations. Each blue point represents the average across 10 turbulent realizations, with error bars indicating the standard deviation.}
    \label{fig:Turb_lines_new}
\end{figure*}

\begin{figure*}
    \centering
	\includegraphics[width=0.9\textwidth]{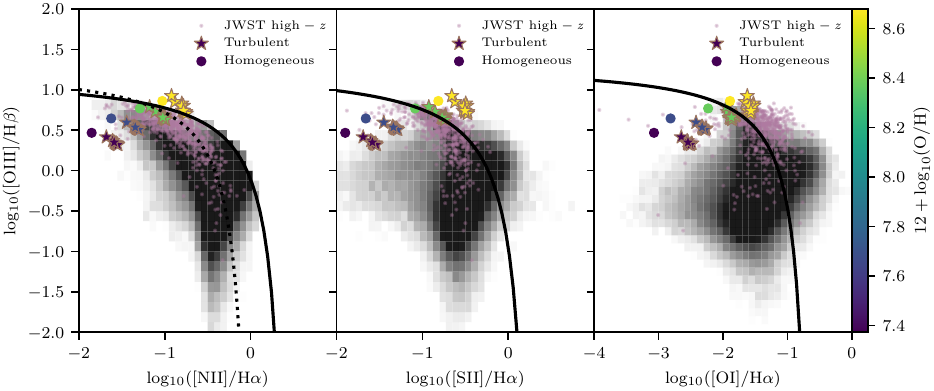}
    \caption{Comparison of diagnostic emission-line ratios from the homogeneous and turbulent simulations with observational data. The left panel shows $\rm log_{10}([N~II]/H\alpha)$ vs. $\rm log_{10}([O~III]/H\beta)$, the center panel shows $\rm log_{10}([S~II]/H\alpha)$ vs. $\rm log_{10}([O~III]/H\beta)$, and the right panel shows $\rm log_{10}([O~I]/H\alpha)$ vs. $\rm log_{10}([O~III]/H\beta)$. Small purple dots represent high redshift galaxies from the DAWN JWST Archive \citep{BrammerValentino2025}. Homogeneous simulations are the larger dots and turbulent simulations are the brown-outlined stars. Both are colored by metallicity. The background gray histogram indicates the number density of galaxies from the Sloan Digital Sky Survey (SDSS, \citealt{Aihara2011}). The solid lines in each panel represent relations from \protect\cite{Kewley:2002} to separate star-forming galaxies/H~II regions from AGN. The dashed line in the first panel is the empirical discriminator \protect\cite{Kauffmann_2003}. AGN are filtered out according to the \protect\cite{Kewley:2002} relation for $\rm log_{10}([N~II]/H\alpha)$ vs. $\rm log_{10}([O~III]/H\beta)$.}
    \label{fig:Kewley_Jin}
\end{figure*}

\begin{figure}
	\includegraphics{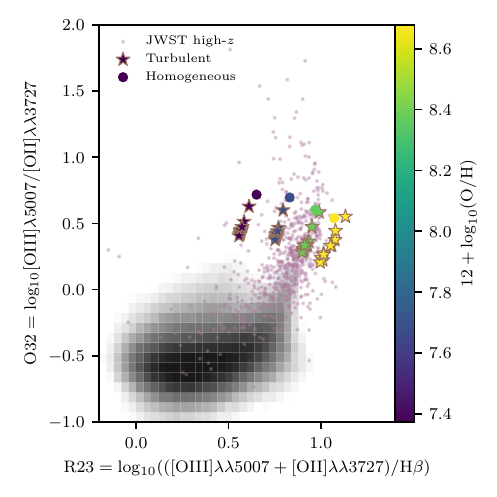}
    \caption{Comparison of O32 versus R23 from the homogeneous and turbulent simulations with observational data. Note that [O~III]$~\rm \lambda\lambda5007$ denotes the O~III nebular doublet consisting of [O~III]~$\rm \lambda4959$ and [O~III]~$\rm \lambda5007$, and [O~II]~$\rm \lambda\lambda3727$ denotes the O~II doublet consisting of [O~II]~$\rm \lambda3726$ and [O~II]~$\rm \lambda3729$. Purple dots represent high redshift galaxies from the DAWN JWST Archive \citep{BrammerValentino2025}. Homogeneous simulations are represented by large circles and turbulent simulations are the brown-outlined stars, with each colored by metallicity. The background gray histogram indicates the number density of galaxies from the Sloan Digital Sky Survey (SDSS, \citealt{Aihara2011}).}
    \label{fig:R23_O32}
\end{figure}

In Figure~\ref{fig:turb} we show maps of a density slice, a column density map, a temperature slice, an integrated O~{\small II} map, an  O~{\small II} slice, and an integrated O~{\small III} map for one of the turbulent realizations. Within the slice, the density can vary by four orders of magnitude while the temperature varies by up to $\sim5,000$~K. The shape of the H~{\small II} region is complex and is traced by O~{\small II}, while O~{\small III} fills the medium by volume. These properties are representative of all of the turbulent simulations, independent of metallicity. 

\cite{Jin:2022b} recently showed that in a turbulent medium, emission lines that trace the edge of the ionization fronts (i.e. [N~{\small II}] and [O~{\small II}]) tend to increase in luminosity compared to a homogeneous medium of the same mean density while volume filling emission lines (e.g. [O~{\small III}]) tend to decrease. Our simulations show the exact same behaviour. In Figure~\ref{fig:Turb_lines_new} we show the ratios of [O~{\small III}]/H$\beta$, [N~{\small II}]/H$\alpha$, and [O~{\small III}]/[O~{\small II}] for the constant density simulations (orange points) compared with the turbulent realizations (blue points with error bars). We see systematic differences between the homogeneous simulations and the turbulent volumes at the same mean density where [O~{\small III}]/H$\beta$ decreases, [N~{\small II}]/H$\alpha$ increases and [O~{\small III}]/[O~{\small II}] decreases, consistent with \cite{Jin:2022b}. These shifts are not insignificant, in the case of [O~{\small III}]/[O~{\small II}], the change can be as much as a factor of two while for the other ratios, the variation can be up to $\sim40\%$. The exact amount that a line ratio varies is sensitive to the type of turbulence (both mach number and compressive to solenoidal ratio), and this will be explored in future work.

As \cite{Jin:2022b} point out, the likely physical origin of the enhancement in emission from boundary species is due to the surface area of the ionization front being enhanced by the complex density field. Likewise, most of the nebula by volume is at lower density than the mean which may suppress emission from the volume filling species.

Stated more rigorously, the luminosity of an emission line is proportional to both density and volume such that $L\propto n_{\rm ion}n_eV$, where $n_{\rm ion}$ is the number density of the relevant ion (e.g. O$^+$, O$^{++}$, N$^+$, etc.), $n_e$ is the electron number density, and $V$ is the volume of the H~{\small II} region. In the case of pure hydrogen, in a homogeneous medium, the density dependence divides out and the luminosity of the Balmer lines goes as $L\propto Q$, where $Q$ is the emission rate of ionizing photons per second. In principle, the hydrogen recombination lines act as a photon counter and the same should be true in the case of the turbulent H~{\small II} regions, which is why we see little variation in their luminosities. However, the same is not true for the other strong collisionally excited metal lines that we consider. The strength ionization parameter at the Strömgren radius partially dictates the size of the transition layer between ionized and neutral gas. If the surface area of the H~{\small II} region increases, simply by conservation of photons, the average ionization parameter at the transition layer must decrease. Lowering the ionization parameter increases the size of the transition layer which would increase the emission from the boundary species. For large enough density contrasts, the H~{\small II} region should preferentially terminate in higher density gas which would also decrease $U$.

Our results have clear importance for interpreting observations because [O~{\small III}]/H$\beta$ is commonly used as a metallicity diagnostic, the combination of [O~{\small III}]/H$\beta$ and [N~{\small II}]/H$\alpha$ can differentiate between different types of ionization sources, while [O~{\small III}]/[O~{\small II}] is considered to trace ionization parameter. Systematic changes in these line ratios due to complex density structure would bias the interpretation of observed spectra. This is demonstrated in Figure~\ref{fig:Kewley_Jin} where we show our homogeneous and turbulent models on the classic BPT \citep{Baldwin1981} and VO \citep{Veilleux1987} diagrams with respect to local SDSS galaxies and high-redshift JWST observations. Indeed, the turbulent models, for a fixed mean density and metallicity, can span a significant region of parameter space. 

Additionally, we see that turbulence causes a non-negligible shift in how we interpret the theoretical calibrations for total excitation and ionization parameter. This is shown in Figure \ref{fig:R23_O32}, where we compare R23, an indicator of total excitation, to O32, a tracer for ionization parameter \citep[e.g.][]{Kewley:2019}. We see that turbulence causes up to a 0.4~dex decrease in O32 and up to a 0.1~dex decrease in R23. Hence, a highly turbulent medium would typically be inferred to have lower ionization and excitation. 

Having shown that the emission line ratios fundamentally change due to geometry, below we consider how a complex turbulent geometry impacts inferences of density, temperature, and metallicity.

\section{Inferring Density, Temperature, and Metallicity from Turbulent H~{\small II} Regions}
\label{sec:results}

\subsection{Temperature}
Before we can infer the metallicity of the turbulent H~{\small II} regions, it is crucial to consider the variety of electron temperatures that can be used in $T_{e}$-based metallicity measurements. Following the standard two-zone approximation \citep{1992AJ....103.1330G}, we assume separate electron temperatures for the boundary-tracing O~{\small II} zone and volume-filling O~{\small III} zone. To derive $T_{\rm O~{\small III}}$, we adopt the direct method approach of utilizing the ratio temperature of [O~{\small III}]~$\lambda$4363/$\lambda$5007 \citep{1941ApJ....93..230M}. For $T_{\rm O~{\small II}}$, one of two separate approaches is generally employed in literature. When accurate measurements of O~{\small II} auroral lines are available, $T_{\rm O~{\small II}}$ is measured using the O~{\small II} auroral lines. Otherwise, it is derived by assuming a theoretically or empirically determined relation between $T_{\rm O~{\small II}}$ and $T_{\rm O~{\small III}}$ \citep[e.g.][]{Campbell1986, 1992AJ....103.1330G}. To account for the different approaches, we experiment with four separate estimation schemes of $T_{\rm O~{\small II}}$. 
\\ \\
1. $T_{\rm O~{\small II}}$ = $T_{\rm O~{\small III}}$: an approximation that effectively adopts a one-zone approach. 
\\ \\
2. The direct $T_{\rm O~{\small II}}$ method using $R_{\lambda\lambda3727}$: the typical approach of estimating $T_{\rm O~{\small II}}$ to be the ratio temperature of the O~{\small II} auroral quadruplet to the strong O~{\small II} doublet \citep{2003MNRAS.346..105P}.
\\ \\
3. The direct $T_{\rm O~{\small II}}$ method using only [O~{\small II}]~$\lambda3726$: nearly identical to method two, except only considering the strong O~{\small II} line with the higher critical density. In doing so, we aim to reduce the effect of density inhomogeneities on the line ratio.
\\ \\
4. $T_{\rm O~{\small II}} = 0.7T_{\rm O~{\small III}}+0.3$: a widely adopted theoretical relation from \cite{Campbell1986}, based on models of \cite{Stasiska1982}.
\\ \\
In addition, we repeat the use of Method 2 to infer $T_e$ and metallicity for the homogeneous simulations, establishing a baseline of comparison for the effects of turbulence.

Figure \ref{fig:Line_Temps} shows that across all turbulent realizations, our inferred values of $T_{\rm O~{\small II} }$ are systematically higher than the line temperature. The magnitude of deviation and scatter increase with lower metallicity / higher $T_{\rm O~{\small III}}$. Correspondingly, our inferred $T_{\rm O~{\small II}}$~-~$T_{\rm O~{\small III}}$ relations are much steeper than the relation between the O~{\small II} line temperature and $T_{\rm O~{\small III}}$.

\subsection{Metallicity}
With $T_{\rm O~{\small II} }$ and $T_{\rm O~{\small III}}$ determined, we compute direct-method metallicities using Equation~\ref{eq:metallicity}.
We adopt various $T_{\rm O~{\small II} }$ to calculate the emissivity ratio of H$\beta$ and [O~{\small II}]~$\lambda\lambda3727$ and the direct $T_{\rm O~{\small III}}$ to calculate the emissivity ratio of H$\beta$ and [O~{\small III}]~$\lambda5007$. We note that electron density also affects the value of emissivity and therefore adopt the density value measured from the ${R}_{{\rm O~II}}$.

Figure~\ref{fig:T_e} shows that across all turbulent realizations, our inferred metallicities deviate from the true values by $0.02-0.1$~dex, with typical offsets closer to 0.05~dex. The scatter increases with metallicity. Different assumptions on $T_{\rm O~{\small II}}$ introduce $<0.05$~dex of scatter into the inferred metallicity, which is a smaller effect than the presence of turbulence. The metallicites we infer from the turbulent realizations are always systematically below those we infer from the homogeneous realizations, which differ from true metallicity values by $<0.02$~dex regardless of metallicity. This implies that turbulence does indeed systematically bias direct method metallicity estimates.

Of the four approaches of estimating $T_{\rm OII}$, the \cite{Campbell1986} relation yields the most accurate estimates at all metallicities. For low-metallicity realizations, the second and third variations ($R_{\lambda\lambda3727}$ and $R_{\lambda3726}$) outperform the simple $T_{\rm O~{\small II} } = T_{\rm O~{\small III}}$ approximation, but the opposite trend holds at high metallicites. Furthermore, using $R_{\lambda3726}$ consistently yields more accurate results, although the difference is $<0.01$~dex and generally not statistically significant.

\begin{figure*}
    \includegraphics{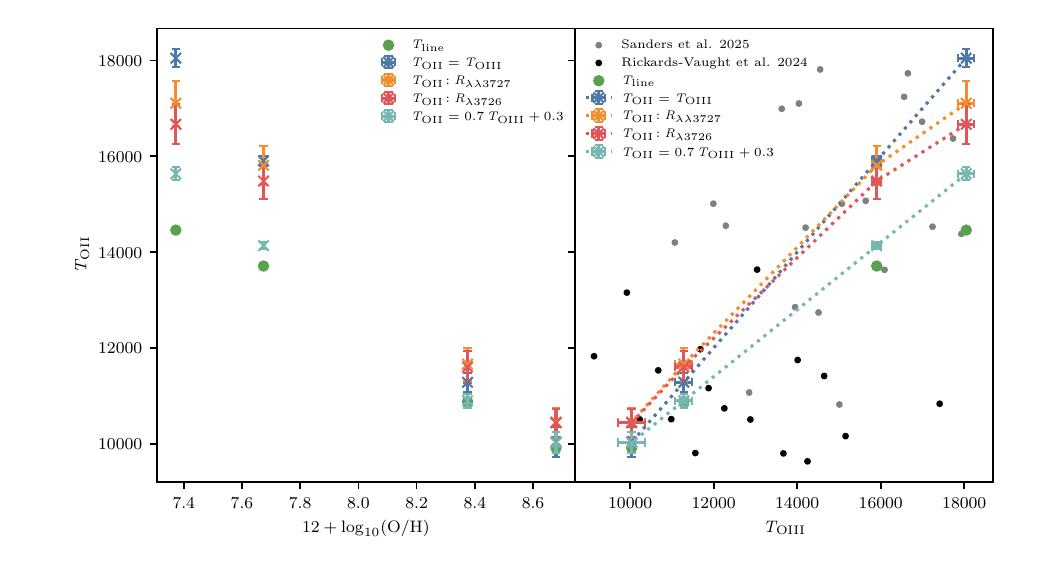}
    \caption{(Left) Comparison of $T_{\rm O~II }$ values measured from four different approaches of the direct method to the [O~II] line temperature defined by \cite{Stasinska:1978}. Green dots represent the line temperature at each metallicity, and the colored $\times-$marks represent the values of $T_{\rm O~II}$ obtained using the different approaches. Each turbulent data point represents the mean of measured metallicities from 10 realizations of the same metallicity, with error bars indicating the standard deviation. (Right) Comparison of $T_{\rm O~II}$ and $T_{\rm O~III }$ for turbulent simulations at 4 metallicities. Here, $T_{\rm O~III}$ is measured using the $R_{\lambda4363}$ line ratio and $T_{\rm O~II}$ is measured using the 4 methods outlined in Section \ref{sec:results} and plotted as $\times$-marks that are color coded accordingly. The  O~{II} line temperatures as defined in Equation \ref{eq:line_temp} of simulations of each metallicity are plotted as green dots. Each turbulent data point represents the mean of measured metallicities from 10 realizations of the same metallicity, with error bars indicating the standard deviation. We provide comparisons with similarly-derived observational data from local H~II regions \citep{vaught2024investigatingdriverselectrontemperature} (black dots) and high redshift JWST data \citep{2025arXiv250810099S} (gray dots).}
    \label{fig:Line_Temps}          
\end{figure*}

\begin{figure}
    \includegraphics{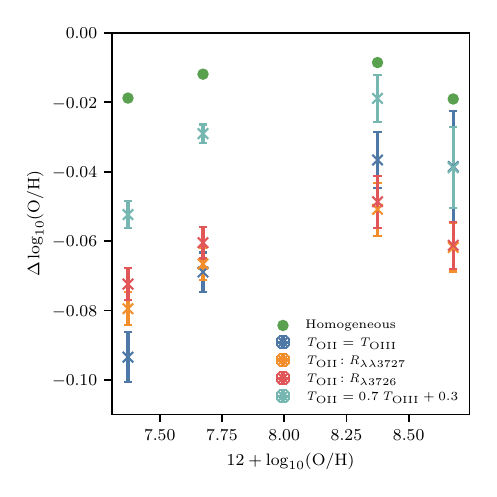}
    \caption{Comparison of the offset between the true metallicity of simulations and measured metallicities. Metallicity was measured for the homogeneous and turbulent realizations, with the green dots representing the homogeneous simulations and the colored $\times-$marks representing the turbulent realizations. Four different approaches were taken to measure metallicity for the turbulent realizations, with the only difference being the method of estimating the value of $T_{\rm O~{\small II}}$ to be utilized in Equation \ref{eq:metallicity}. Each turbulent data point represents the mean of measured metallicities from 10 realizations of the same metallicity, with error bars indicating the standard deviation.}
    \label{fig:T_e}
\end{figure}

However, it is worth noting that the choice of $T_{\rm O~{\small II} }$ affects our metallicity estimates by at most 0.05~dex, likely due to the fact that the overall mass of O$^{++}$ is greater than that of O$^{+}$ by a factor of $\sim2 - 3$ (depending on metallicity). As seen in Figure~\ref{fig:Line_Temps}, our estimates of $T_{\rm O~{\small II} }$ can differ with respect to each other by $>2,000~{\rm K}$, and yet the estimates of metallicity still remain within 0.05~dex of each other. For lower ionization parameters, we can expect the choice of $T_{\rm O~{\small II}}$ to have a much more significant effect on the direct method estimate of metallicity. The choice of how one calculates $T_{\rm O~{\small II}}$ is therefore likely more important in the lower redshift Universe where ionization parameters tend to be lower.

\begin{figure}
    \includegraphics{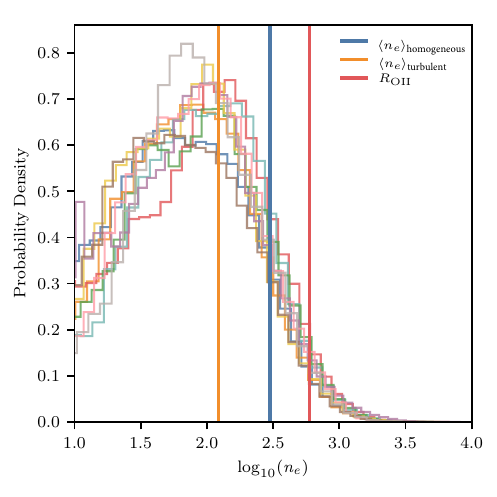}
    \includegraphics{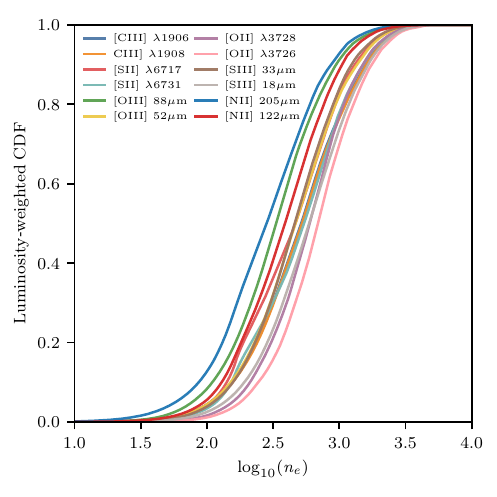}
    \caption{(Top) Electron density PDF of the 0.05~$Z_{\odot}$ simulations. The volumetric mean density of homogeneous simulations, mean volumetric mean density of turbulent simulations, and the $R_{\mathrm{OII}}$ line-ratio derived density are labeled using blue, orange, and red vertical lines, respectively. (Bottom) Cumulative distribution functions of emission line luminosity as a function of electron density for all lines used in this work for the same 0.05~$Z_{\odot}$ model averaged over the ten realizations. Here we see that the luminosity is always dominated by gas above the mean volumetric density.}
    \label{fig:density_hist}          
\end{figure}

\subsection{Density}
Having quantified the discrepancy between $T_e$-derived metallicities and the true metallicity of the simulated H~{\small II} regions, we now consider the electron density. However, for the case of electron density, the question of which ``true'' electron density we should recover is nontrivial. For photoionization models simulated at constant density, such as our homogeneous models, the electron density remains constant in the fully ionized regions\footnote{Note that it can decrease in partially ionized gas.}. Alternatively, both empirically and theoretically, one may expect density stratification in a nebula \citep{Stasinska:1978,1992AJ....103.1330G,Wang_2004}. In the top panel of Figure \ref{fig:density_hist} we show an electron density histogram of the 0.05~$Z_{\odot}$ turbulent simulations. We see that the volumetric mean density of the homogeneous simulations is $\sim300$~cm$^{-3}$, while the volumetric mean density of the turbulent simulations is $\sim100$~cm$^{-3}$. Additionally, we see that turbulence causes electron density to scatter between $10-500$~cm$^{-3}$ and extend to densities $~3000$~cm$^{-3}$, confirming the presence of density variations. As a result of this, we expect density diagnostics relying on different parent ions to probe different parts of the H~{\small II} region and provide different values of electron density. Taking this into account, we compare the values of electron density measured from our calibrated density diagnostics with the mass-weighted density of the parent ion as well as the luminosity-weighted densities of the emission lines used in the diagnostic. In doing so, we find that neither the mass-weighted densities of the parent ions nor the initial/final volumetric mean density of the H~{\small II} region are accurately probed by the density diagnostics. 

\begin{figure*}
    \centering
    \includegraphics{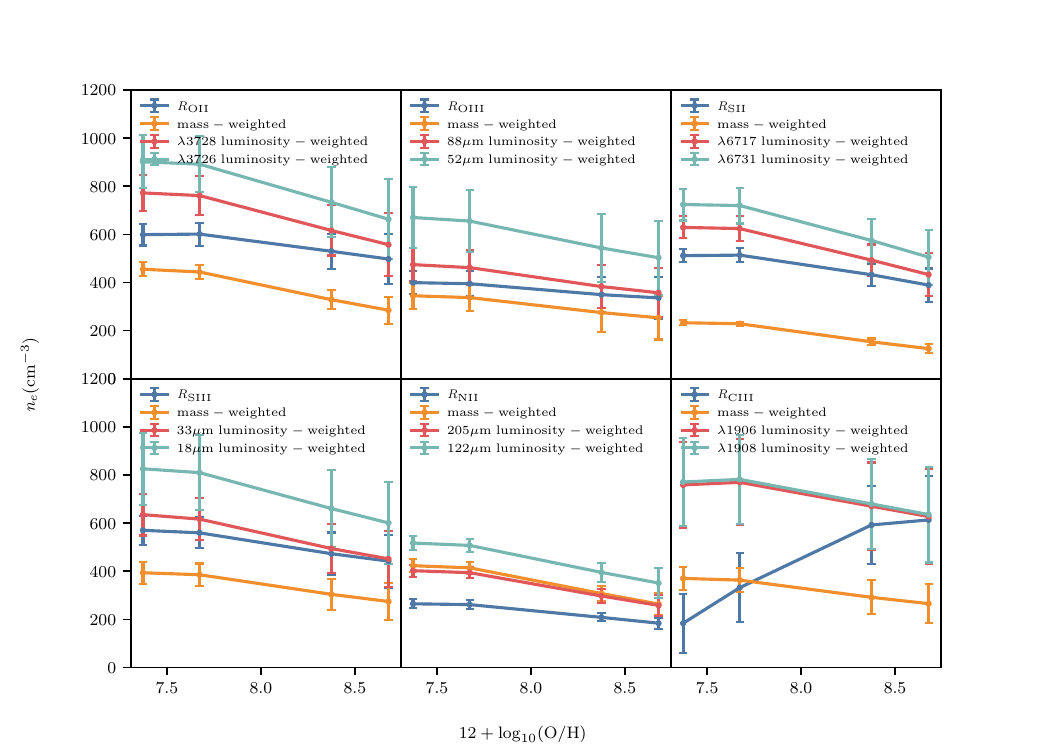}
        \caption{Comparison of emission-line-derived densities (blue) using ${R}_{{\rm O~II}}$, ${R}_{{\rm O~III}}$, ${R}_{{\rm S~II}}$, ${R}_{{\rm S~III}}$, ${R}_{{\rm N~II}}$, ${R}_{{\rm C~III}}$ at different metallicities. Additionally, each emission-line-derived metallicity is plotted against the mass-weighted electron density (orange) of the ion corresponding to the line ratio and the luminosity-weighted electron densities (red and cyan, respectively) of the emission lines used to construct the line ratio.}
    \label{fig:density_comparison}          
\end{figure*}

\begin{figure*}
    \centering
    \includegraphics{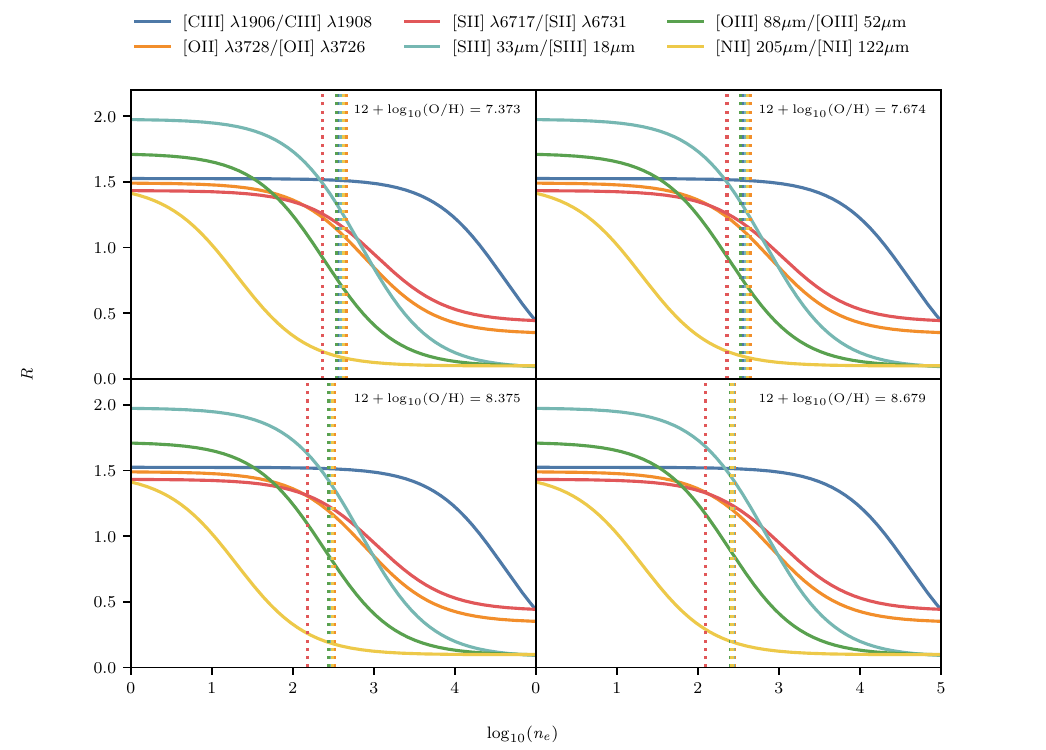}
    \caption{Line-ratio as a function of electron density for every density-sensitive diagnostic used in this work. For each metallicity, the mass-weighted electron densities of each ion (displayed in Figure \ref{fig:density_comparison}) are marked with dotted vertical lines color coded with the same color of the line ratio diagnostic constructed using the corresponding parent ion.}
    \label{fig:Kewley}          
\end{figure*}

As shown in Figure~\ref{fig:density_comparison}, all diagnostics except ${R}_{{\rm N~II}}$ systematically overestimate the mass-weighted densities. Typical offsets range from 100 to 300 cm$^{-3}$ ($\sim30-50\%$ higher), reflecting the fact that line emission is disproportionately biased toward the denser parts of the H~{\small II} regions. In contrast, the diagnostics are generally in agreement (to within 100~cm$^{-3}$) with the luminosity-weighted density of the emission line with the lower critical density, with the exception of ${R}_{{\rm O~II}}$, ${R}_{{\rm N~II}}$, and ${R}_{{\rm C~III}}$. We note that two of these exceptions can be excluded for physical reasons. ${R}_{{\rm N~II}}$ consistently overestimates the mass-weighted densities and ${R}_{{\rm C~III}}$ shows no consistent trend and significant scatter. Both effects can be traced to the fact that their mass-weighted densities fall outside the density-sensitive regimes of the line-ratios (Figure~\ref{fig:Kewley}), making them highly sensitive to small fluctuations. Furthermore, although variations in the temperature used to calibrate density diagnostics (typically $10^4$~K) alters the density-sensitive regimes of line-ratios by negligible amounts, their effect on the density-insensitive regimes of line-ratios are significant. We assume a uniform temperature of $10^4$~K for all of our temperature diagnostics, which inevitably deviates from the electron temperature characteristic of each ionization zone. As such, the trends we find that pertain to ${R}_{{\rm N~II}}$ and ${R}_{{\rm C~III}}$ are not reflective of the trends that should pertain to density diagnostics.

\section{Discussion}
\label{sec:discussion}
In this section, we discuss the underlying physics behind the trends found in Section \ref{sec:results}. In addition, we contextualize the magnitude of the effects of turbulence in relation to the relevant literature and identify future research directions.
\\
\\
\subsection{Abundance Discrepancy Problem}
Several studies have shown gas-phase metallicities derived from the direct $T_e$ method are systematically lower than those derived from recombination line (RL) methods by $\geq0.2$~dex, the abundance discrepancy problem  \citep{Esteban_2004,Garcia_Rojas_2007}. Our simulations indicate that even in the presence of supersonic turbulence, the direct method recovers true metallicity values to within 0.1~dex, as long as the mach number remains below $\sim5.5$. We verified that applying RL methods to our simulations result in more accurate metallicity measurements than direct method-derived metallicites; as such, turbulence alone does not seem to account for the entire abundance discrepancy factor (ADF). 

Several explanations have been proposed for the ADF, including temperature fluctuations \citep[e.g.][]{1967ApJ...150..825P,Esteban_2002,Peimbert_2013}, chemical inhomogeneities \citep{Yuan_2010}, and non-Maxwellian electron distributions such as $\kappa$ distributions \citep{Nicholls_2012}. Nevertheless, our results show that the complex internal structures driven by turbulence do in fact result in systematic underestimates in direct method metallicities. And although our simulations do not fully recover the ADF found in observational studies, they confirm the role turbulence plays in affecting metallicity and electron temperature diagnostics.

\subsection{Discrepancy between $T_{\rm O~{\small II} }$ Approaches}
We found that inferring $T_{\rm O~{\small II}}$ using $R_{\lambda\lambda3727}$, which in theory should be more accurate, does not outperform the theoretical \cite{Campbell1986} relation, and for higher metallicities, also does not outperform the one-zone approximation. As seen in Figure \ref{fig:Line_Temps}, the relation of $T_{\rm O~{\small II} }~-~T_{\rm O~{\small III} }$ we obtain while using $R_{\lambda\lambda3727}$ coincides far more closely to the simplistic one-zone approximation as opposed to the relation between the line temperatures of $T_{\rm O~{\small II} }$ and $T_{\rm O~{\small III} }$. This finding is surprising; however, we offer some potential explanations for this phenomena.

Firstly, H~{\small II} regions have been known to exhibit isothermal conditions at certain metallicites. For example, in the models of \cite{Kewley:2019} they find that the isothermal conditions apply at a metallicity of $\rm 12+\log(O/H)\sim 8.23$. Thus, it is expected for the one-zone approximation to work well for H~{\small II} regions with select metallicites, though this can vary depending on the exact conditions. In Figure \ref{fig:Line_Temps}, we observe this metallicity to be $\rm 12+\log(O/H)\sim 8.7$ for our simulations. In contrast, as the metallicity of H~{\small II} regions deviate significantly from 8.7, a stronger temperature gradient forms, leading to inaccuracies in the one-zone approximation. This is the result of two key physical mechanisms. First, higher metallicity leads to increased radiative cooling. This is particularly true in the regions dominated by O$^{++}$ due to the efficiency of fine structure cooling. Second, the higher abundance of metals causes increased radiation hardening towards the outer edges of the H~{\small II} region, which leads to stronger heating of the regions dominated by O$^{+}$ \citep{Kewley:2019}. The balance between these two effects determines when isothermal conditions are reached. This clearly depends on metallicity, density, and the shape of the radiation field.

Furthermore, studies have found that the density sensitive [O~{\small II}]~$\lambda\lambda3727$ doublet causes the $R_{\lambda\lambda3727}$ line ratio to be affected by density inhomogeneities by a non-negligible amount \citep{vaught2024investigatingdriverselectrontemperature}. This leads to over-estimations of $T_{\rm O~{\small II}}$, which ultimately result in the systematic under-prediction of metallicity, coinciding with the finding of studies such as \cite{choustikov2025megatrondisentanglingphysicalprocesses}. At high redshifts, a similar process likely impacts $T_{\rm O~{\small III}}$ \citep{Martinez2025}. As shown in the previous section, removing the higher critical density [O~{\small II}]~$\lambda3729$ line does reduce the effects of density inhomogeneities by a marginal amount. The density used to calibrate the $R_{\lambda\lambda3727}$ line ratio is itself not entirely accurate, as it is computed using ${R}_{{\rm O~II}}$, which we have determined earlier to be overly biased towards the high density regions. All of these factors contribute to the systematic error in the estimation of $T_{\rm O~{\small II} }$, ultimately leading to the consistent metallicity underestimate of $\sim0.06$~dex.

Our results suggest that theoretical $T_{\rm O~{\small II} }$~-~$T_{\rm O~{\small III}}$ relations may outperform methods that utilize the O~{\small II} auroral lines as a result of the density dependence of [O~{\small II}]~$\lambda\lambda3727$ (see Figure \ref{fig:Line_Temps}).

\subsection{Physical Interpretation of Inferred Densities in Turbulent H~{\small II} Regions}

Several studies have shown that applying constant density diagnostics to H~{\small II} regions with complex density structures will recover values that are characteristic of the regions producing the most emission line strength as opposed to a volumetric-mean of the entire H~{\small II} region \citep{Kewley:2019}. Our results in the previous section are consistent with this finding, as we have shown that the electron density estimates inferred from the turbulent H~{\small II} regions better reflect select luminosity-weighted densities as opposed to mass or volume-weighted densities. Specifically, we found that the line ratio densities align with the luminosity-weighted density of the lower critical density emission line of each density diagnostic. 

The dominance of high density regions in line emission can be readily inferred from Figure \ref{fig:density_hist}. As mentioned in the previous section, the electron densities within the simulations extend to values as high as $\sim3000$~cm$^{-3}$. This high density tail is important because the emission line luminosity of a parcel of gas scales as $n^2$ (until the critical density of the line is reached). Thus even though the high density gas represents a small fraction of the mass and volume, it can contribute significantly to emission line luminosity. For example, although we typically have $\sim13-16\times$ more gas cells with an electron density of $~100\ {\rm cm^{-3}}$ compared to   $\gtrsim1000\ {\rm cm^{-3}}$, the factor of ten difference in density in principle results in a factor of $100$ difference in luminosity, which more than accounts for the relative sparsity of higher density gas parcels. There is nuance for individual lines in that as density changes, the ionization states may differ and once the critical density is reached, luminosity increases as $n$ rather than $n^2$. To explore this further, in the bottom panel of Figure~\ref{fig:density_hist} we show cumulative distribution functions of the luminosity of individual emission lines as a function of electron density averaged over the ten realizations of the $0.05~Z_{\odot}$ model. Indeed we find that the luminosity is always dominated by gas significantly denser the mean volumetric density.

\subsection{Discrepancy between Density Diagnostics}
As described in the previous sections, studies have shown that density diagnostics relying on different parent ions probe different ionization zones within H~{\small II} regions and thus recover different values of electron density \citep{Wang_2004,2023MNRAS.523.2952M}. Figure~6 of \citealt{choustikov2025megatrondisentanglingphysicalprocesses} has shown that this effect is very pronounced on galactic scales using cosmological simulations with a multi-phase ISM. However, it remains unclear how much the measured densities of different density diagnostics should differ from each other on the scale of individual H~{\small II} regions. Nevertheless, we showed in previous sections that our inferred densities from diagnostics of different parent ions differ with each other by $\leq 200$~cm$^{-3}$. As above, this result is certainly sensitive to the properties of the turbulence, which motivates running more simulations to systematically study a larger parameter space.

\subsection{Caveats}
\label{sec:caveats}
Our turbulent H~{\small II} regions are undoubtedly simplistic --- they represent a first step beyond simple slab or spherical geometries that are nearly ubiquitous among published photoionization models. Our 3D setup is likely more physical than simple geometries, but still misses key physical effects that impact real H~{\small II} regions. For example, the ISM is known to be turbulent \citep[e.g.][]{Larson1981,Elmegreen2004}, but the exact nature (compressive to solenoidal ratio, driving scale, power spectrum, mach number etc.) likely differs between environments. Transsonic and subsonic turbulence will produce emission line ratios more similar to the spherical models \citep{Gray:2017}. We have specifically chosen a numerical setup where turbulence has an impact on the emission line ratios; however, our chosen parameters are unlikely to be representative of the ISM across different environments. A more systematic study varying the properties of the turbulence is warranted. Ideally one would even be able to infer the turbulence from the emission line ratios and their line shapes, but this is beyond the scope of our current work.

Arguably the most important assumption of our work is the exclusion of hydrodynamics. We placed a single O star at the center of numerous realizations of a turbulent gas density field and evolved them to ionization and thermal equilibrium. This assumption may be overly simplistic under two circumstances. First if the density is too low ($\lesssim1~{\rm cm^{-3}}$), the timescale for the ionization front to reach the Strömgren radius becomes comparable than the lifetime of a massive O star. Implications of this are discussed in \textcolor{blue}{Katz et al. {\it in prep.}}; however, in the context of this work, the box size is not large enough to capture this effect and thus the ionizing radiation leaks from the volume (i.e. $f_{\rm esc}>0$). Since $f_{\rm esc}$ is typically $<10\%$, our results are not strongly impacted by leaking radiation. Second, in real H~{\small II} regions, the gas is not static, but rather is sloshing around from turbulence, impacted by gravity, and subject to feedback from radiation pressure, photoheating, and stellar winds \citep[e.g.][]{Spitzer1978,Draine2011,Klessen2016}. These effects are particularly important when the ionization front transitions from R-type to D-type. Moreover, depending on the medium, the radiation pressure or pressure from photoheating may circularize the H~{\small II} region and drive a shell \citep[e.g.][]{Draine2011b}. In this case, a spherical H~{\small II} region may be more appropriate than what we have assumed in our models. All of these effects motivate the need for simulations that self-consistently model the physics of star formation and stellar feedback in high-resolution, turbulent, magnetized molecular clouds (see e.g. \citealt{Kimm2019,Grudic2021,Chon2021,Kimm2022,Bate2023,Menon2025}). Work towards this end with {\small RAMSES-RTZ} is currently underway.

\section{Conclusions}
\label{sec:conclusions}
We have presented a suite of simulations at various metallicities designed to study the impact of density inhomogeneities driven by supersonic turbulence on density, temperature, and metallicity inferences from emission line ratios in static H~{\small II} regions. Our work represents a step beyond the simple geometries of prior photoionization models, an assessment of how the physical condition of turbulence affects the efficacy of current nebular emission diagnostics. Our main results and conclusions are as follows:
\begin{enumerate}
    \item Supersonic turbulence can drive significant systematic changes in strong emission line ratios when compared to homogeneous models of the same mean density. More specifically, we found [O~{\small III}]/H$\beta$ decreases, [N~{\small II}]/H$\alpha$ increases, and [O~{\small III}]/[O~{\small II}] decreases, consistent with the fractal models of \cite{Jin:2022b}.
    \item We compute the metallicities of homogeneous and turbulent simulations using multiple variations of the direct method (Figure \ref{fig:T_e}). We show that turbulence leads to systematic underestimates of metallicity $<0.1$~dex. This deficit is less than the observationally determined abundance discrepancy factor of $>0.2$~dex \citep{Esteban_2004,Garcia_Rojas_2007}.
    \item We compare the values of metallicity and $T_{\rm O~{\small II} }$ determined using different variations of the direct method (Figure~\ref{fig:Line_Temps}, \ref{fig:T_e}). We find that the density inhomogeneities caused by turbulence affect methods that utilize the O~{\small II} doublet to a larger extent than methods that utilize a theoretically determined $T_{\rm O~{\small II}}-T_{\rm O~{\small III}}$ relation.
    \item We compare the electron density diagnostics to the mass-weighted electron density of the corresponding parent ion and luminosity-weighted electron density of the corresponding emission lines (Figure~\ref{fig:density_comparison}). In agreement with previous studies \citep{Kewley:2019}, we find that diagnostic-derived electron densities are a poor tracer of the volumetric mean electron density of an H~{\small II} region or the mass-weighted electron density of any given ion. They instead align closely with the luminosity-weighted density of the lower critical density line used in the diagnostic.
\end{enumerate}

Our numerical implementation of turbulence, remains an important first step towards the development of more complex, realistic photo-ionization models. The magnitude by which turbulence can impact observed emission line ratios motivates the need for more complex models of H~{\small II} regions.

\section*{Acknowledgments}

We thank the referee for their constructive comments on the manuscript. The authors thank Romain Teyssier for both developing and open-sourcing {\small RAMSES}.

This work made extensive use of the dp265 and dp016 projects on the DiRAC ecosystem. HK is particularly grateful to Christopher Mountford and Alastair Basden for support on DIaL3 and Cosma8, respectively. HK is especially thankful for the support on Glamdring provided by Jonathan Patterson. The material in this manuscript is based upon work supported by NASA under award No. 80NSSC25K7009. 
AJC gratefully acknowledges support from the Cosmic Dawn Center through the DAWN Fellowship. The Cosmic Dawn Center (DAWN) is funded by the Danish National Research Foundation under grant No. 140.

This work used the DiRAC@Durham facility managed by the Institute for Computational Cosmology on behalf of the STFC DiRAC HPC Facility (\url{www.dirac.ac.uk}). The equipment was funded by BEIS capital funding via STFC capital grants {\small ST/P002293/1}, {\small ST/R002371/1} and {\small ST/S002502/1}, Durham University and STFC operations grant {\small ST/R000832/1}. This work also used the DiRAC Data Intensive service at Leicester, operated by the University of Leicester IT Services, which forms part of the STFC DiRAC HPC Facility. The equipment was funded by BEIS capital funding via STFC capital grants {\small ST/K000373/1} and {\small ST/R002363/1} and STFC DiRAC Operations grant {\small ST/R001014/1}. DiRAC is part of the National e-Infrastructure.

\bibliographystyle{mn2e}
\bibliography{References}


\end{document}